\documentclass[aps,prd,onecolumn,preprintnumbers,groupedaddress,showpacs,nofootinbib,amssymb]{revtex4-2}
\usepackage{graphicx,color}
\usepackage{amsmath}
\usepackage{amssymb}
\usepackage{amsfonts}
\usepackage{bm}
\usepackage{cancel}
\usepackage{slashed}

\newcommand{\e}{\mathrm{e}}

\allowdisplaybreaks[4]
\begin{document}

\preprint{KEK-TH-2786, KEK-Cosmo-0404}
\title{Apparent Phantom Crossing in Gauss-Bonnet Gravity}
\author{Shin'ichi~Nojiri,$^{1,2}$}
\email{nojiri@nagoya-u.jp}
\author{Sergei~D.~Odintsov,$^{3,4}$}
\email{odintsov@ieec.cat}
\author{V.~K.~Oikonomou,$^{5,6}$}
\email{v.k.oikonomou1979@gmail.com;voikonomou@gapps.auth.gr}
\affiliation{$^{1)}$ Theory Center, High Energy Accelerator Research Organization (KEK), \\
Oho 1-1, Tsukuba, Ibaraki 305-0801, Japan \\
$^{2)}$ Kobayashi-Maskawa Institute for the Origin of Particles
and the Universe, Nagoya University, Nagoya 464-8602, Japan \\
$^{3)}$ ICREA, Passeig Luis Companys, 23, 08010 Barcelona, Spain\\
$^{4)}$ Institute of Space Sciences (IEEC-CSIC) C. Can Magrans
s/n, 08193 Barcelona, Spain\\
$^{5)}$ Department of Physics, Aristotle University of
Thessaloniki, Thessaloniki 54124, Greece\\
$^{6)}$Center for Theoretical Physics, Khazar University, 41
Mehseti Str., Baku, AZ-1096, Azerbaijan}

\begin{abstract}
The recent observations of the Dark Energy Spectroscopic Instrument (DESI) indicated the possibility that the dark energy equation of state parameter $w$ might change from $w<-1$ to $w>-1$ when the redshift $z\sim 0.5$, which is called the inverse phantom crossing.
In this paper, we investigate the possibility of the phantom crossing, and we construct realistic models realizing the crossing in the framework of the scalar--Einstein--Gauss-Bonnet gravity and ghost-free $f(\mathcal{G})$ gravity.
We also investigate the scenario of the apparent phantom crossing, where dark matter energy density decreases more slowly than usually expected, which might explain the DESI observations.
In the scenarios developed, the energy conditions are not violated by any component of the cosmic fluid.
In the framework of the apparent phantom crossing, we also propose a new scenario, where the particle corresponding to the scalar field in the scalar--Einstein--Gauss-Bonnet gravity is dark matter.
The mass of the particle might increase due to the coupling with the Gauss-Bonnet invariant, which makes the decrease of the dark matter energy density slower.
This last scenario may suggest that the inverse phantom crossing might be related to the transition from the decelerating expansion of the Universe to the accelerating expansion.
\end{abstract}

\maketitle

\section{Introduction}\label{SecI}

The accelerated expansion of the present Universe could be generated by an unknown fluid filling the Universe, which is dubbed dark energy.
If the dark energy is a perfect fluid, it is characterized by the ratio $w$ of the pressure $p$ and the energy density $\rho$ of the dark energy fluid, $w=\frac{p}{\rho}$, which is called the equation of state (EoS) parameter.
The cosmological constant corresponds to $w=-1$.
The dark energy with $-\frac{1}{3}>w>-1$ is called quintessence and that with $w<-1$ is phantom.
The quintessence dark energy can be realized by a canonical scalar field with a potential, which could be an analogy with the inflaton, which generates the inflation in the early Universe.
Recently, the observations of the Dark Energy Spectroscopic Instrument (DESI) \cite{DESI:2024mwx, DESI:2025zgx} seem to indicate that there could have been a transition from the phantom Universe to the non-phantom Universe.
The transition is called the phantom crossing~\cite{Cortes:2024lgw, Colgain:2024xqj, Giare:2024smz, Shlivko:2024llw}.
For a mainstream of articles related to the DESI data, see Refs.~\cite{Odintsov:2024woi, Dai:2020rfo, He:2020zns, Nakai:2020oit, DiValentino:2020naf, Agrawal:2019dlm, Ye:2020btb, Vagnozzi:2021tjv, Desmond:2019ygn, Hogas:2023pjz, OColgain:2018czj, Vagnozzi:2019ezj, Krishnan:2020obg, Colgain:2019joh, Vagnozzi:2021gjh, Lee:2022cyh, Krishnan:2021dyb, Ye:2021iwa, Ye:2022afu, Verde:2019ivm, Menci:2024rbq, Adil:2023ara, Reeves:2022aoi, Ferlito:2022mok, Vagnozzi:2021quy, DiValentino:2020evt, Sabogal:2024yha, DiValentino:2025sru, Odintsov:2025kyw, vanderWesthuizen:2025iam, Paliathanasis:2025dcr, Odintsov:2025jfq, Kessler:2025kju, Cai:2025mas, Cheng:2025lod, Nojiri:2025low}.
Formally, the phantom crossing means the transition from the non-phantom Universe to the phantom Universe.
Therefore, we may call this transition inverse phantom crossing.
The observation of DESI seems to indicate that the EoS parameter $w$ could change depending on time.
Furthermore, the phantom dark energy cannot be realized by the canonical scalar field without resorting to tachyon fields.
Therefore, we need a model of dark energy distinct from the canonical scalar field framework.

In this paper, we investigate the scenario of the inverse phantom crossing by using the scalar--Einstein--Gauss-Bonnet gravity and the ghost-free $f(\mathcal{G})$ gravity.
The action of the scalar--Einstein--Gauss-Bonnet gravity is given by~\cite{Nojiri:2005vv, Nojiri:2006je},
\begin{align}
\label{SEGB1}
S=\int d^4 x \sqrt{-g}\left[ \frac{R}{2\kappa^2} - \frac{1}{2}\partial_\mu \phi \partial^\mu \phi - V(\phi) - \xi(\phi) \mathcal{G}\right]\, .
\end{align}
Here the Gauss-Bonnet invariant $\mathcal{G}$ is defined as,
\begin{align}
\mathcal{G} \equiv R^2 - 4 R_{\mu\nu} R^{\mu\nu} + R_{\mu\nu\rho\sigma} R^{\mu\nu\rho\sigma}\, .
\label{GBterm}
\end{align}
In \eqref{SEGB1}, $\phi$ is a scalar field, $V(\phi)$ is the function of $\phi$, and $\xi(\phi)$ is the Gauss-Bonnet invariant coefficient, which is a function of the scalar field $\phi$.

In the canonical scalar field theory with $\xi(\phi)=0$, the expansion of the Universe filled with a perfect fluid with the equation of state parameter $w>-1$ can be realized, but this is impossible in the case that the EoS of the Universe is $w<-1$.
Here, the EoS parameter $w$ is the ratio of the pressure $p$ and the energy density $\rho$ of the perfect fluid, $w\equiv \frac{p}{\rho}$.
In the framework of the scalar field with $\xi(\phi)=0$, the Universe corresponding to $w<-1$ can be realized only by a non-canonical scalar field, where the signature before the kinetic term $\frac{1}{2}\partial_\mu \phi \partial^\mu \phi$ is not minus but plus, $-\frac{1}{2}\partial_\mu \phi \partial^\mu \phi\to + \frac{1}{2}\partial_\mu \phi \partial^\mu \phi$.
Such a scalar field is, however, a ghost, which is physically inconsistent.
As a classical theory, the kinetic energy of ghosts is unbounded below.
On the other hand, in the quantum theory, the ghosts generate negative norm states, which is not consistent with quantum theory, although the Fadeev-Popov ghosts in the gauge theories could be well-known and tractable~\cite{Kugo:1977zq, Kugo:1979gm}, but the Fadeev-Popov ghosts appear in the combinations of zero norm.

In \cite{Nojiri:2005vv}, however, it has been shown that we can realize the Universe corresponding to $w<-1$ without resorting to ghost fields, in the framework of the scalar--Einstein--Gauss-Bonnet gravity, by choosing,
\begin{align}
\label{GBssk}
V(\phi) \propto \e^{-\frac{2\phi}{\phi_0}}\, , \quad \xi(\phi) \propto \e^{\frac{2\phi}{\phi_0}}\, .
\end{align}
Here $\phi_0$ is a constant and $\xi(\phi)$ is negative.
In \cite{Nojiri:2023mvi}, the unification of inflation with early and late dark energy was also proposed by using the scalar--Einstein--Gauss-Bonnet gravity.

The observation of GW170817, which is a gravitational wave generated by the merger of a black hole and a neutron star, gives a constraint on the time dependence of $\xi(\phi)$.
The GW170817 event gives the following constraint on the propagating speed $c_\mathrm{GW}$ of the gravitational wave as follows,
\begin{align}
\label{GWp9}
\left| \frac{{c_\mathrm{GW}}^2}{c^2} - 1 \right| < 6 \times 10^{-15}\, ,
\end{align}
where $c$ denotes the speed of light.
This tells us that the propagation speed of the gravitational wave must almost coincide with the speed of light.
The constraint requires~\cite{Nojiri:2023jtf},
\begin{align}
\label{condition}
\nabla_\mu \nabla^\nu \xi = \frac{1}{4}g_{\mu\nu} \nabla^2 \xi \, .
\end{align}
We may consider the Friedmann-Lema\^{i}tre-Robertson-Walker (FLRW) Universe, whose metric is given by the following line element,
\begin{align}
\label{FRW}
ds^2= -dt^2 + a(t)^2\sum_{i=1,2,3} \left(dx^i\right)^2\, .
\end{align}
Here $t$ is the cosmological time and $a(t)$ is the scale factor.
Then Eq.~\eqref{condition} becomes a second-order ordinary differential equation with respect to the cosmological time $t$ as follows~\cite{Odintsov:2020xji},
\begin{align}
\label{FLRWcond}
\ddot \xi = H \dot \xi \, ,
\end{align}
whose solution is
\begin{align}
\label{FLRWsolsol}
\dot \xi = \xi_0 + \xi_1 \int dt a(t)\, ,
\end{align}
which gives the time-dependence of the coefficient function $\xi(\phi)$.
We should note, however, that the condition \eqref{condition} cannot be satisfied in a general spacetime, like a black hole spacetime~\cite{Nojiri:2023jtf}, except for a trivial case that $\xi(\phi)$ is a constant.
We should also note that the observed oldest event of the merger of a black hole and a neutron star occurred when the redshift { $z\sim 0.04$}, which could be about { one} billion years ago depending on models~\cite{LIGOScientific:2021qlt}.
Therefore, there is no constraint like \eqref{GWp9} of the time-dependence of $\xi(\phi)$ as in \eqref{condition} before one billion years ago.
For example, the (inverse) phantom crossing observed in the DESI observation could have occurred about 5-8 billion years ago { corresponding to $z\sim 0.5$}.
It is possible that the coefficient function $\xi(\phi)$ may have drastically evolved and played a dominant role in the early Universe.

The $f(\mathcal{G})$ gravity model~\cite{Nojiri:2005jg, Cognola:2006eg, Leith:2007bu, Li:2007jm, Kofinas:2014owa, Zhou:2009cy} was considered to eliminate the scalar field $\phi$, as a dynamical degree of freedom, in the scalar--Einstein--Gauss-Bonnet gravity~\eqref{SEGB1}.
The $f(\mathcal{G})$ gravity model has, however, ghosts~\cite{DeFelice:2009ak}.
In order to eliminate the ghost, the model with a constraint given by the Lagrange multiplier field was proposed~\cite{Nojiri:2018ouv, Nojiri:2022cah}.
This model is the ghost-free $f(G)$ gravity.

Although the phenomenon of the phantom crossing is still not established, and it might be an artifact of the choice in parametrization, in this paper, we construct realistic models showing the inverse phantom crossing in the framework of the scalar--Einstein--Gauss-Bonnet gravity and the ghost-free $f(G)$ gravity\footnote{
For the model of the dark energy in the framework of the scalar--Einstein--Gauss-Bonnet gravity from the point view of particle physics, see \cite{Hussain:2025vbo, Arora:2025ecj} although the phantom crossing was not found there.
}.
We also construct models where the standard phantom crossing and the inverse phantom crossing iterate alternately.
Recently, in \cite{Khoury:2025txd}, a new scenario, which might solve the problem in the DESI observations, was proposed.
In this scenario, if the energy density of the dark matter decreases more slowly than $a^{-3}$, the DESI observations might be explained without the real transition from $w<-1$ to $w>-1$ of the EoS parameter of dark energy.
We consider this scenario by using the scalar--Einstein--Gauss-Bonnet gravity and the ghost-free $f(G)$ gravity.
In this scenario, the total energy density and the total pressure do not violate the energy conditions; the components included in the total energy and pressure can satisfy the energy conditions.
Furthermore, we consider the scenario by assuming that the particles corresponding to the scalar field $\phi$ in the action \eqref{SEGB1} become dark matter.
Due to the coupling $\xi(\phi)$ with the Gauss-Bonnet invariant $\mathcal{G}$ in \eqref{SEGB1}, the mass depends on the curvature and by properly choosing the potential $V(\phi)$ and the coupling $\xi(\phi)$, the energy density can decrease slower than $a^{-3}$.
This last scenario also suggests that the inverse phantom crossing might occur due to the transition from the decelerating expansion of the Universe to the accelerating expansion.

\section{
Scalar--Einstein--Gauss-Bonnet gravity}\label{SecII}

We start with the gravitational action of the so-called scalar--Einstein--Gauss-Bonnet gravity~\cite{Nojiri:2005vv, Nojiri:2006je} in \eqref{SEGB1}.
The field equations obtained by the variation of the action \eqref{SEGB1} with respect to the metric $g_{\mu\nu}$ and to the scalar field $\phi$ have the following forms,
\begin{align}
\label{gb4b}
\frac{1}{2\kappa^2}\left( R^{\mu\nu} - \frac{1}{2} g^{\mu\nu} R\right)
=&\, T^{\mu \nu} + T_\mathcal{G}^{\mu \nu} \, , \nonumber \\
T_\mathcal{G}^{\mu \nu}\equiv &\, \left(\frac{1}{2} \partial^\mu \phi \partial^\nu \phi
 - \frac{1}{4}g^{\mu\nu} \partial_\rho \phi \partial^\rho \phi \right) - \frac{1}{2} g^{\mu\nu} V(\phi) \nonumber \\
&\, + 2 \left( \nabla^\mu \nabla^\nu \xi(\phi)\right)R - 2 g^{\mu\nu} \left( \nabla^2\xi(\phi)\right)R
 - 4 \left( \nabla_\rho \nabla^\mu \xi(\phi)\right)R^{\nu\rho}
 - 4 \left( \nabla_\rho \nabla^\nu \xi(\phi)\right)R^{\mu\rho} \nonumber \\
&\, + 4 \left( \nabla^2 \xi(\phi) \right)R^{\mu\nu}
+ 4g^{\mu\nu} \left( \nabla_{\rho} \nabla_\sigma \xi(\phi) \right) R^{\rho\sigma}
 - 4 \left(\nabla_\rho \nabla_\sigma \xi(\phi) \right) R^{\mu\rho\nu\sigma}\, , \\
\label{scalarGB1}
0=&\, \Box \phi - V'(\phi) - \xi'(\phi) \mathcal{G}\, ,
\end{align}
where $T^{\mu \nu}$ is the energy-momentum tensor for the matter fields.

We consider a flat FLRW metric \eqref{FRW} so the $(00)$, $(ij)$ components of the field equations~\eqref{gb4b} and the scalar field equation \eqref{scalarGB1} can be written as,
\begin{align}
\label{SEGB3}
\frac{3}{\kappa^2}H^2 =&\, \rho_\mathcal{G} + \rho \, , \quad
\rho_\mathcal{G} \equiv \frac{1}{2}{\dot\phi}^2 + V(\phi) + 24 H^3 \frac{d \xi(\phi(t))}{dt} \, ,\nonumber \\
 - \frac{1}{\kappa^2}\left(2\dot H + 3 H^2 \right) =&\, p_\mathcal{G} + p \, , \quad
p_\mathcal{G} \equiv \frac{1}{2}{\dot\phi}^2 - V(\phi)
 - 8H^2 \frac{d^2 \xi(\phi(t))}{dt^2} - 16H \dot H \frac{d\xi(\phi(t))}{dt} - 16 H^3 \frac{d \xi (\phi(t))}{dt} \, , \\
\label{scalarGB1BB}
0=&\, - \ddot \phi - 3 H \dot \phi - V'(\phi)
- 24 H^2 \left( H^2 + \dot H \right) { \xi'(\phi)}\, ,
\end{align}
where the Hubble rate $H$ is defined by $H\equiv \frac{\dot a}{a}$ and $\rho$ and $p$ are the energy density and pressure of the matter fields, respectively.
The energy density $\rho$ and pressure $p$ include all the contributions from matter content of the Universe with an equation of state parameter $w_m$,
\begin{align}
\label{SGBEG9_0}
\rho = \sum_m \rho_m \, , \quad p = \sum_m p_m = \sum w_m \rho_m \, , \quad w_m \equiv \frac{p_m}{\rho_m} \, .
\end{align}
We assume that each matter component satisfies the conservation law,
\begin{align}
\label{SGBEG9} 0 = \dot \rho_m + 3H \left( \rho_m + p_m \right) \, ,
\end{align}
so these are perfect fluids.
If $w_m$ is a constant, Eq.~\eqref{SGBEG9} can be easily solved in terms of the scale factor as follows,
\begin{align}
\label{matters} \rho_m = \rho_{m0} a^{-3 \left( 1 + w_m \right)} = \rho_{m0} \e^{-3 \left( 1 + w_m \right) N}\, ,
\end{align}
where $\rho_{m0}$ is a constant and $N$ is the number of $e$-foldings $a=\e^N$.
By using the scalar field equation \eqref{scalarGB1BB} and the Bianchi identities, we can show that $\rho_\mathcal{G}$ and $p_\mathcal{G}$ also satisfy the
conservation law,
\begin{align}
\label{SGBEG9GB}
0 = \dot \rho_\mathcal{G} + 3H \left( \rho_\mathcal{G} + p_\mathcal{G} \right) \, .
\end{align}
This equation is used when we discuss the phantom crossing by regarding $\rho_\mathcal{G}$ and $p_\mathcal{G}$ as contributions from the dark energy.
When the phantom crossing occurs, since we have $\rho_\mathcal{G}=-p_\mathcal{G}$, then $\dot\rho_\mathcal{G}$ must vanish.
In the phantom Universe, we find $\rho_\mathcal{G}<-p_\mathcal{G}$ and therefore $\dot \rho_\mathcal{G}$, and in the non-phantom Universe, $\rho_\mathcal{G}>p_\mathcal{G}$.
Therefore, there is a crossing from the non-phantom Universe to the phantom one if $\ddot\rho_\mathcal{G}>0$, and there is an inverse crossing if $\ddot\rho_\mathcal{G}<0$.

For later use, we rewrite the field equations~\eqref{SEGB3} in terms of $N$ instead of the cosmic time $t$ as follows,
\begin{align}
\label{SEGB3N}
0=&\, - \frac{3}{\kappa^2}H^2 + \frac{1}{2}H^2{\phi'}^2 + V(\phi) + 24 H^4 \frac{d \xi(\phi(N))}{dN}
+ \sum_m \rho_{m0} \e^{-3 \left( 1 + w_m \right) N} \, ,\nonumber \\
0 =&\, \frac{1}{\kappa^2}\left(2H H' + 3 H^2 \right) + \frac{1}{2}H^2 {\phi'}^2 - V(\phi)
 - 8 \left( 2 H^4 \frac{d \xi (\phi(N))}{dN} + 3 H^3 H' \frac{d \xi(\phi(N))}{dN} + H^4 \frac{d^2 \xi(\phi(N))}{dN^2} \right) \nonumber \\
&\, + \sum_m w_m \rho_{m0} \e^{-3 \left( 1 + w_m \right) N} \, ,
\end{align}
where $'=\frac{d}{dN}$ and we have used $\frac{d\phi}{dt}=H\frac{d\phi}{dN}$, and so on.
In terms of $N$, the conservation law of the dark energy can be rewritten as,
\begin{align}
\label{SGBEG9GBN}
0 = {\rho_\mathcal{G}}' + 3 \left( \rho_\mathcal{G} + p_\mathcal{G} \right) \, .
\end{align}
Then, when the phantom crossing occurs, we have ${\rho_\mathcal{G}}'=0$.
A crossing from the non-phantom Universe to the phantom one occurs if ${\rho_\mathcal{G}}''>0$, an inverse crossing if ${\rho_\mathcal{G}}''<0$.

We now review the reconstruction procedure based on Refs.~\cite{Nojiri:2006je}. We eliminate the scalar potential $V(\phi)$ by combining two equations in \eqref{SEGB3N},
\begin{align}\label{GBEDE1}
0=&\, \frac{2H H'}{\kappa^2} + H^2 {\phi'}^2
 - 8 H \e^N \frac{d}{dN} \left( \e^{-N} H^3 \frac{d \xi (\phi(N))}{dN} \right) + \sum_m \left( 1 + w_m\right) \rho_{m0} \e^{-3 \left( 1 + w_m \right) N} \, ,
\end{align}
which can be integrated to obtain $\xi\left(\phi\left(N\right)\right)$ as follows,
\begin{align}
\label{GBEDE2}
\xi\left(\phi\left(N\right)\right) =&\, \frac{1}{8} \int_{N^{(2)}}^N dN_1\frac{\e^{N_1}}{H \left( N_1 \right)^3} \int_{N^{(1)}}^{N_1} dN_2 \e^{-N_2} \nonumber \\
&\, \times \left\{ \frac{2H' \left(N_2 \right)}{\kappa^2} + H\left( N_2 \right) \phi' \left( N_2 \right)^2
+ \frac{1}{H\left(N_2\right)} \sum_m \left( 1 + w_m\right) \rho_{m0} \e^{-3 \left( 1 + w_m \right) N_2} \right\}\, ,
\end{align}
where $N^{(1)}$ and $N^{(2)}$ are integration constants.
By using the first equation in \eqref{SEGB3N}, we find the expression for the scalar potential,
\begin{align}
\label{GBEDE3}
V\left(\phi\left(N\right)\right) =&\, \frac{3}{\kappa^2}H\left(N\right)^2 - \frac{1}{2}H\left(N\right)^2 \phi'\left(N\right)^2
 - 3 H\left(N\right) \e^N \int_{N^{(1)}}^N dN_1 \e^{-N_1} \left\{ \frac{2H' \left(N_1 \right)}{\kappa^2} \right. \nonumber \\
&\, \left. + H\left( N_1 \right) \phi' \left( N_1 \right)^2
+ \frac{1}{H\left(N_1\right)} \sum_m \left( 1 + w_m\right) \rho_{m0} \e^{-3 \left( 1 + w_m \right) N_1} \right\}
 - \sum_m \rho_{m0} \e^{-3 \left( 1 + w_m \right) N} \, .
\end{align}
We may assume a model described by the following scalar potential and coupling in terms of two functions $g=g(N)$ and $f=f(\phi)$,
\begin{align}
\label{GBEDE4}
\xi\left(\phi\right) =&\, \frac{1}{8} \int_{N^{(2)}}^{f(\phi)} dN_1\frac{\e^{N_1}}{g \left( N_1 \right)^3} \int_{N^{(1)}}^{N_1} dN_2 \e^{-N_2} \nonumber \\
&\, \times \left\{ \frac{2g' \left(N_2 \right)}{\kappa^2} + \frac{g \left( N_2 \right)}{f' \left( f^{-1}\left(N_2\right) \right)^2}
+ \frac{1}{g\left(N_2\right)} \sum_m \left( 1 + w_m\right) \rho_{m0} \e^{-3 \left( 1 + w_m \right) N_2} \right\}\, , \\
\label{GBEDE5}
V\left(\phi\right) =&\, \frac{3}{\kappa^2}g\left(f\left(\phi\right)\right)^2 - \frac{g\left(f\left(\phi\right)\right)^2}{2f'\left(\phi \right)^2}
 - 3 g\left(f\left(\phi\right)\right) \e^{f(\phi)} \int_{N^{(1)}}^{f(\phi)} dN_1 \e^{-N_1} \left\{ \frac{2 g' \left(N_1 \right)}{\kappa^2} \right. \nonumber \\
&\, \left. + \frac{g\left( N_1 \right)}{f' \left( f^{-1}\left(N_1\right) \right)^2}
+ \frac{1}{g\left(N_1\right)} \sum_m \left( 1 + w_m\right) \rho_{m0} \e^{-3 \left( 1 + w_m \right) N_1} \right\}
 - \sum_m \rho_{m0} \e^{-3 \left( 1 + w_m \right) f^{-1}(\phi)} \, .
\end{align}
Then the following solutions for the Hubble rate $H$ and the scalar field $\phi$ are obtained,
\begin{align}
\label{GBEDE6}
H(N)=g\left( N \right)\, , \quad \phi=f^{-1}\left(N\right)\, .
\end{align}
Here $f^{-1}(N)$ is the inverse function of $f(N)$.

\section{Model of Inverse Phantom Crossing}\label{SecIII}

\subsection{A Realistic Model}

Based on the formulation in the last section, we consider in this section a realistic model exhibiting inverse phantom crossing, and specifically, we choose,
\begin{align}
\label{ex1}
\rho_\mathcal{G} = \frac{\Lambda_1}{\e^{\alpha \left(N-N_0\right)} + \e^{-\alpha \left(N-N_0\right)}} + \Lambda_0 \, .
\end{align}
Here $\Lambda_1$, $\Lambda_0$, $\alpha$, and $N_0$ are constants and we assume that $\Lambda_1$, $\Lambda_0$ and $\alpha$ are positive.
Since,
\begin{align}
\label{ex1_1}
{\rho_\mathcal{G}}' =&\, - \frac{\Lambda_1\alpha \left( \e^{\alpha \left(N-N_0\right)} - \e^{-\alpha \left(N-N_0\right)}\right)}
{\left( \e^{\alpha \left(N-N_0\right)} + \e^{-\alpha \left(N-N_0\right)}\right)^2} \, , \nonumber \\
{\rho_\mathcal{G}}'' =&\, - \frac{\Lambda_1\alpha^2 \left\{ \left( \e^{\alpha \left(N-N_0\right)} + \e^{-\alpha \left(N-N_0\right)}\right)^2
 - 2 \left( \e^{\alpha \left(N-N_0\right)} - \e^{-\alpha \left(N-N_0\right)}\right)^2 \right\}}
{\left( \e^{\alpha \left(N-N_0\right)} + \e^{-\alpha \left(N-N_0\right)}\right)^3} \nonumber \\
=&\, \frac{\Lambda_1\alpha^2 \left\{ \left( \sqrt{2} + 1 \right) \e^{\alpha \left(N-N_0\right)}
 - \left(\sqrt{2} - 1 \right)\e^{-\alpha \left(N-N_0\right)}\right\}
\left\{ \left( \sqrt{2} - 1 \right) \e^{\alpha \left(N-N_0\right)}
 - \left( \sqrt{2} + 1 \right)\e^{-\alpha \left(N-N_0\right)} \right\}}
{\left( \e^{\alpha \left(N-N_0\right)} + \e^{-\alpha \left(N-N_0\right)}\right)^3} \, ,
\end{align}
we find,
\begin{align}
\label{ex1_2}
{\rho_\mathcal{G}}' \left( N = N_0 \right) = 0 \, , \quad
{\rho_\mathcal{G}}'' \left( N = N_0 \right) = - \frac{\Lambda_1\alpha^2}{2}<0 \, .
\end{align}
This tells that the inverse phantom crossing occurs at $N=N_0$.
We now choose $g(N)$ in \eqref{GBEDE4} and \eqref{GBEDE5} as
\begin{align}
\label{ex1_3}
\frac{3}{\kappa^2} g(N)^2 = \rho_\mathcal{G} + \rho_m = \frac{\Lambda_1}{\e^{\alpha \left(N-N_0\right)} + \e^{-\alpha \left(N-N_0\right)}}
+ \Lambda_0 + \sum_m \rho_{m0} \e^{-3 \left( 1 + w_m \right) N} \, ,
\end{align}
and $f(\phi)$ is arbitrarily chosen, and construct a model by following \eqref{GBEDE4} and \eqref{GBEDE5}, so we find the solution,
\begin{align}
\label{ex1_4}
\frac{3}{\kappa^2} H(N)^2 = \rho_\mathcal{G} + \rho_m = \frac{\Lambda_1}{\e^{\alpha \left(N-N_0\right)}
+ \e^{-\alpha \left(N-N_0\right)}} + \Lambda_0 + \sum_m \rho_{m0} \e^{-3 \left( 1 + w_m \right) N} \, ,
\end{align}
Therefore, we obtain a model that generates the inverse phantom crossing explicitly.
Regarding the matter fluids, we may include radiation and dust, with the latter corresponding to dark matter and baryonic matter,
\begin{align}
\label{ex1_5}
\sum_m \rho_{m0} \e^{-3 \left( 1 + w_m \right) N} = \rho_\mathrm{d} \e^{-3N} + \rho_\mathrm{r} \e^{-4N}\, .
\end{align}
We choose $N=0$ at present, then $\rho_\mathrm{d}$ and $\rho_\mathrm{r}$ are the present values of the energy densities of the dust and the radiation, respectively.
Then $\frac{\rho_\mathrm{r}}{\rho_\mathrm{d}} \sim 2.9656\times 10^{-4}$. When $\alpha=3$, we should be a little bit careful because for large $N$, the first term of the r.h.s. in Eq.~\eqref{ex1_4} or Eq.~\eqref{ex1_3} behaves as,
\begin{align}
\label{ex1_6}
 \frac{\Lambda_1}{\e^{\alpha \left(N-N_0\right)} + \e^{-\alpha \left(N-N_0\right)}} \sim \Lambda_1 \e^{3N_0} \e^{-3N}\, ,
\end{align}
as in the case of dust, and therefore, the density of the first term may shift the energy density of the dust effectively.
If $\alpha \gg 3$, the first term may be neglected in the present Universe.

The phantom crossing could occur about 5-8 billion years ago, that is, the redshift $z$ had the value $z\sim 0.5$, which gives $N_0\sim -0.4$.
Although the problem of the Hubble tension (see \cite{Riess:2021jrx}, for example), which is the discrepancy of the value of the Hubble constant in the present Universe between the observed values from the early Universe~\cite{Planck:2018vyg} (at the recombination epoch) and the local Universe~\cite{Riess:2020fzl}, we may assume the present value of $H$ as $H_0 \sim 70\, \mathrm{m}/\left(\mathrm{s}\cdot\mathrm{MPc}\right)$.
Then Eq.~\eqref{ex1_4} gives,
\begin{align}
\label{ex1_7}
\frac{3}{\kappa^2} {H_0}^2 = \frac{\Lambda_1}{\e^{- \alpha N_0} + \e^{\alpha N_0}} + \Lambda_0 + \rho_\mathrm{d} + \rho_\mathrm{r} \, .
\end{align}
On the phantom crossing $N=N_0<0$, $H$ could be $H = H_\mathrm{pc} \sim 80-90\, \mathrm{m}/\left(\mathrm{s}\cdot\mathrm{MPc}\right)$, and Eq.~\eqref{ex1_4} gives,
\begin{align}
\label{ex1_8}
\frac{3}{\kappa^2} {H_\mathrm{pc}}^2 = \frac{\Lambda_1}{2} + \Lambda_0 + \rho_\mathrm{d} \e^{-3N_0} + \rho_\mathrm{r} \e^{-4N_0} \, .
\end{align}
For a given $\alpha$, we may solve Eqs.~\eqref{ex1_7} and \eqref{ex1_8} with respect to $\Lambda_0$ and $\Lambda_1$ and obtain a more realistic model.


The data obtained from the observations of CMB, BAO, etc., have been used to determine the evolution of the Hubble rate $H$ \cite{Simon:2004tf, Stern:2009ep, Moresco:2012jh, Blake:2012pj, Zhang:2012mp, BOSS:2013igd, BOSS:2014hwf, Moresco:2015cya, Moresco:2016mzx, BOSS:2016wmc}.
We may be able to compare the data of the CMB, BAO, etc., with the quantities obtained from the behavior of $H$ in our model, but it could be logically better to directly compare the behavior of $H$ in our model with the evolution of $H$ obtained from the observed data.
In this section, we adjusted the parameters by using the values of $H$ at the present Universe and in the epoch of the inverse phantom crossing as in \eqref{ex1_7} and \eqref{ex1_8}.
Therefore, the values of $H$ in our model should be inside the error bars obtained from the observations of CMB, BAO, etc., and therefore, the model could be consistent.
Furthermore,  before the phantom crossing, the extra term, that is, the first term in \eqref{ex1_3}, is exponentially small and can be neglected.

{
Eq.~\eqref{ex1} tells that when $\alpha\left(N-N_0\right)\to \pm\infty$, $\rho_\mathcal{G} \to \Lambda_0$. 
THis tells $V(\phi)\to \Lambda_0$ and $\xi(\phi)\to \mathrm{constant}$ when $\alpha\left(N-N_0\right) \to \pm\infty$ in the action \eqref{SEGB1} 
and therefore the constraint \eqref{GWp9} coming from the gravitational wave could be satisfied in the late universe. 
As we mentioned, the observed oldest event of the merger of a black hole and a neutron star corresponds to the redshift $z\sim 0.04$. 
On the other hand, the phantom crossing observed DESI corresponds to $z\sim 0.5$ and therefore $\e^{N_0} = \frac{1}{1 + 0.5} \sim 0.67$. 
Let the number of the $e$-foldings corresponding to the oldest event of the merger of a black hole and a neutron star be $N_\mathrm{mrgr}$. 
Then we find $\e^{N_\mathrm{mrgr}} = \frac{1}{1+0.04} \sim 0.96$ and we find $\e^{\alpha \left(N_\mathrm{mrgr} - N_0\right)} \sim \left(1.44\right)^\alpha$. 
Therefore, if we choose $\alpha$ large enough, the constraint \eqref{GWp9} could be satisfied

We may also consider the constraints coming from the Big Bang Nucleosynthesis (BBN) as in \cite{Asimakis:2021yct}, where some constraints have obtained by considering the shift of the freeze-out temperature in the higher derivative gravities from that in the $\Lambda$CDM model. 
The BBN occurred when the redshift is given by $z\sim 10^8-10^9$, that is, when $\e^{N-N_0}\sim 10^{-8}-10^{-9}$. 
Then, as clear from the expression of \eqref{ex1}, the shift from the $\Lambda$CDM model is very small in the epoch of the BBN, and therefore the model \eqref{ex1} does not suffer the constraints from the BBN. 
The shift from the $\Lambda$CDM model appears only at the epoch of the phantom crossing, suggesting that the differences in structure formation, etc., from those in the $\Lambda$CDM model may also be exponentially suppressed. 
See also \cite{Yang:2024kdo, Yang:2025mws} for works that constrain higher-derivative gravity theories from DESI observations.

}

\subsection{An Oscillating Model}

It might be possible that the standard phantom crossing and the inverse crossing often happened in the early Universe.
An old question is why the Universe started the accelerating expansion about 5-8 billion years ago.
Where does the scale of the 5-8 billion years come from?
The scale is not the scale of particle physics or gravity, either.
Of course, the scale is similar to the scale of the age of the Universe, 13.8 billion years.
If the standard phantom crossing and the inverse crossing iteratively occurred and the period becomes longer and longer due to the age of the Universe, the scale of 5-8 billion years might be natural.

As a toy model, instead of \eqref{ex1}, we may consider,
\begin{align}
\label{itcrssng}
\rho_\mathcal{G} = \tilde\Lambda_1 \cos^2 \left( \tilde\alpha \ln \left( N - \tilde N_0 \right) \right) + \tilde \Lambda_0 \, .
\end{align}
Here $\tilde\Lambda_1$, $\tilde\Lambda_0$, $\tilde\alpha$, and $\tilde N_0$ are constants and we assume that $\tilde\Lambda_1$, $\tilde\Lambda_0$, and $\tilde\alpha$ are positive.
When $\tilde\alpha \ln \left( N - \tilde N_0 \right) = \frac{n\pi}{2}$ with an integer $n$, or
\begin{align}
\label{trit}
N=\tilde N_0 \e^\frac{n\pi}{2\tilde\alpha} \, ,
\end{align}
we find ${\rho_\mathcal{G}}'=0$ and therefore transitions between phantom dark energy and non-phantom dark energy occur.
Eq.~(\ref{trit}) indicates that the interval in $N$ of the transitions becomes larger and larger when $n$ becomes larger.
When $n$ is an even number, that is, $n=2m$ with another integer $m$, we obtain ${\rho_\mathrm{G}}''<0$ and therefore the transition corresponds to the inverse phantom transition.
On the other hand, when $n$ is an odd number, $n=2m+1$, we find ${\rho_\mathrm{G}}''>0$ and therefore the transition corresponds to the standard phantom transition, that is the transition from non-phantom dark energy to phantom dark energy.


\section{Phantom Crossing in Ghost-free $f(\mathcal{G})$ Gravity}\label{SecIV}

In the scalar--Einstein--Gauss-Bonnet gravity~\eqref{SEGB1}, the propagation of the scalar field $\phi$ might generate the fifth
force, and the fluctuation may affect the structure formation in the Universe.
In order to avoid such problems, an easy way could be to eliminate the scalar field.
The $f(\mathcal{G})$ gravity { model}~\cite{Nojiri:2005jg, Cognola:2006eg, Leith:2007bu, Li:2007jm, Kofinas:2014owa, Zhou:2009cy} was one of such candidates.

The $f(\mathcal{G})$ gravity model is a counterpart theory of the scalar-Einstein-Gauss-Bonnet gravity model~\cite{Nojiri:2005jg,
Cognola:2006eg, Leith:2007bu, Li:2007jm, Kofinas:2014owa, Zhou:2009cy}, and the action is given by,
\begin{equation}
\label{GB1b}
S=\int d^4x\sqrt{-g} \left(\frac{1}{2\kappa^2}R + f(\mathcal{G}) + \mathcal{L}_\mathrm{matter}\right)\, .
\end{equation}
In the model, however, it has been shown that there ghost degrees of freedom occur \cite{DeFelice:2009ak}.

The ghosts can be eliminated by a constraint given by the Lagrange multiplier field $\lambda$~\cite{Nojiri:2018ouv, Nojiri:2022cah}.
The model is given by the following action,
\begin{align}
\label{gfEGB4}
S_{\mathrm{GB}\phi} = \int d^4 x \sqrt{-g} &\, \left\{ \frac{R}{2\kappa^2}
+ \lambda \left( \omega(\phi) \partial_\mu \phi \partial^\mu \phi - \mu^4 \right) \right. \nonumber \\
&\, \left. - \frac{1}{2} \partial_\mu \phi \partial^\mu \phi - V (\phi) - \xi(\phi) \mathcal{G}
+ \mathcal{L}_\mathrm{matter} \right\}\, ,
\end{align}
Here $\omega(\phi)$ is a function of $\phi$ and $\mu$ is a parameter with a mass dimension.
By varying the action \eqref{gfEGB4} with respect to $\lambda$, the following constraint is obtained,
\begin{align}
\label{gfEGB3}
0 = \omega(\phi) \partial_\mu \phi \partial^\mu \phi - \mu^4 \, .
\end{align}
Due to this constraint, the scalar field $\phi$ becomes non-dynamical and any ghost does not appear in the model \eqref{gfEGB4}~\cite{Nojiri:2018ouv}.
When the signature of the metric is fixed, the function $\omega(\phi)$ is absorbed into the redefinition of the scalar field, but when the signature is changed, as for example at the black hole horizon, the function $\omega(\phi)$ makes the value of the scalar field continuous at the horizon~\cite{Nojiri:2022cah}.

By the variation of the action \eqref{gfEGB4} with respect to the metric $g_{\mu\nu}$,  the following equation corresponding to the Einstein equation is given,
\begin{align}
\label{gb4bD4one}
0= &\, \frac{1}{2\kappa^2}\left(- R_{\mu\nu} + \frac{1}{2} g_{\mu\nu} R\right)
+ \frac{1}{2} g_{\mu\nu} \left\{ \lambda \left( \omega(\phi) \partial_\mu \phi \partial^\mu \phi - \mu^4 \right)
 - \frac{1}{2} \partial_\rho \phi \partial^\rho \phi - V (\phi)\right\} \nonumber \\
&\, - \lambda \omega(\phi) \partial_\mu \phi \partial_\nu \phi
+ \frac{1}{2} \partial_\mu \phi \partial_\nu \phi \nonumber \\
&\, - 2 \left( \nabla_\mu \nabla_\nu \xi(\phi)\right)R
+ 2 g_{\mu\nu} \left( \nabla^2 \xi(\phi)\right)R
+ 4 \left( \nabla_\rho \nabla_\mu \xi(\phi)\right)R_\nu^{\ \rho}
+ 4 \left( \nabla_\rho \nabla_\nu \xi(\phi)\right)R_\mu^{\ \rho} \nonumber \\
&\, - 4 \left( \nabla^2 \xi(\phi) \right)R_{\mu\nu}
 - 4g_{\mu\nu} \left( \nabla_\rho \nabla_\sigma \xi(\phi) \right) R^{\rho\sigma}
+ 4 \left(\nabla^\rho \nabla^\sigma \xi(\phi) \right) R_{\mu\rho\nu\sigma}
+ \frac{1}{2} T_{\mu\nu} \, .
\end{align}
On the other hand, if we vary the action with respect to $\phi$, the following field equation is obtained,
\begin{align}
\label{I10one}
0 =&\, \lambda \omega'(\phi) \partial_\mu \phi \partial^\mu \phi
 - 2 \nabla^\mu \left( \lambda \omega \left(\phi \right) \partial_\mu \phi \right)
+ \nabla^\mu \partial_\mu \phi - V' - \xi' \mathcal{G} \, .
\end{align}
In \eqref{gb4bD4one}, $T_{\mu\nu}$ is the energy-momentum tensor of the matter fluid in \eqref{gb4b}.
We should note that Eq.~\eqref{I10one} can be obtained by using \eqref{gb4bD4one} and the conservation law, and therefore, we do not use Eq.~\eqref{I10one} in the following.

In the FLRW Universe \eqref{FRW}, Eq.~\eqref{gb4bD4one} gives the equations corresponding to the FLRW equations in the following forms,
\begin{align}
\label{SEGB3B}
0=&\, - \frac{3}{\kappa^2}H^2 + \left( - 2 \lambda \omega(\phi) + \frac{1}{2}\right){\dot\phi}^2 + V(\phi)
+ 24 H^3 \frac{d \xi(\phi(t))}{dt}
+ \sum_m \rho_{m0} \e^{-3 \left( 1 + w_m \right) N} \, ,\nonumber \\
0=&\, \frac{1}{\kappa^2}\left(2\dot H + 3 H^2 \right) + \frac{1}{2}{\dot\phi}^2 - V(\phi)
 - 8H^2 \frac{d^2 \xi(\phi(t))}{dt^2}
 - 16H \dot H \frac{d\xi(\phi(t))}{dt} - 16 H^3 \frac{d \xi(\phi(t))}{dt} \nonumber \\
&\, + \sum_m w_m \rho_{m0} \e^{-3 \left( 1 + w_m \right) N} \, .
\end{align}
The constraint \eqref{gfEGB3} has the following form in the FLRW Universe \eqref{FRW}.
\begin{align}
\label{gfEGB3FLRW}
0= \omega(\phi) {\dot\phi}^2 + 1 \, ,
\end{align}
which we used in \eqref{SEGB3B}

As in \eqref{SEGB3N}, we rewrite Eq.~\eqref{SEGB3B} in terms of the $e$-foldings number $N$,
\begin{align}
\label{SEGB3NB}
0=&\, - \frac{3}{\kappa^2}H^2 + \left( - 2 \lambda \omega(\phi) + \frac{1}{2}\right) H^2 \phi'(N)^2
+ V(\phi) + 24 H^4 \frac{d \xi(\phi(N))}{dN} + \sum_m \rho_{m0} \e^{-3 \left( 1 + w_m \right) N} \, ,\nonumber \\
0=&\, \frac{1}{\kappa^2}\left(2H H' + 3 H^2 \right) + \frac{1}{2}H^2 \phi'(N)^2 - V(\phi)
 - 8H^4 \frac{d^2 \xi(\phi(t))}{dN^2}
 - 24 H^3 \frac{dH}{dN} \frac{d\xi(\phi(N))}{dN} \nonumber \\
&\, - 16 H^4 \frac{d \xi(\phi(t))}{dN} + \sum_m w_m \rho_{m0} \e^{-3 \left( 1 + w_m \right) N} \, ,
\end{align}
and by eliminating $V(\phi)$ in \eqref{SEGB3NB}, we obtain,
\begin{align}
\label{GBeq1}
0=&\, \frac{2}{\kappa^2} H(N) H'(N) + \left( - 2 \lambda \omega(\phi) + 1 \right) H(N)^2 \phi'(N)^2
 - 8 H(N)^4 \frac{d^2 \xi(\phi(t))}{dN^2} \nonumber \\
&\, - 24 H(N)^3 \frac{dH}{dN} \frac{d\xi(\phi(N))}{dN} +8 H(N)^4 \frac{d \xi(\phi(t))}{dN}
+ \sum_m \left( 1 + w_m\right) \rho_{m0} \e^{-3 \left( 1 + w_m \right) N} \, .
\end{align}
The equations in \eqref{GBeq1} can be integrated with respect to $\xi(N)$ and take the form,
\begin{align}
\label{SEGB10}
\xi(\phi(N))&= \frac{1}{8}\int^N dN_1 \frac{\e^{N_1}}{H(N_1)^3}
\int^{N_1} \frac{dN_2}{H(N_2) \e^{N_2}} \nonumber \\
&\, \times \left(\frac{2}{\kappa^2}H(N_2) H' (N_2) + \left( - 2 \lambda \left(N_2\right) \omega\left( \phi \left(N_2\right) \right)
+ 1 \right) H(N_2)^2 {\phi'(N_2)}^2
+ \sum_m \left( 1 + w_m\right) \rho_{m0} \e^{-3 \left( 1 + w_m \right) N_2} \right)\, .
\end{align}
Substituting Eq.~\eqref{SEGB10} into the first equation in \eqref{SEGB3NB}, we obtain,
\begin{align}
\label{SEGB11}
V(\phi(N)) &\,
= \frac{3}{\kappa^2}H(N)^2 - \left( - 2 \lambda (N) \omega(\phi) + \frac{1}{2}\right) H(N)^2 \phi' (N)^2
 - \sum_m \rho_{m0} \e^{-3 \left( 1 + w_m \right) N} - 3\e^N H(N) \int^N \frac{dN_1}{H(N_1) \e^{N_1}} \nonumber \\
\times &\,
\left(\frac{2}{\kappa^2} H(N_1) H' (N_1) + \left( - 2 \lambda \left(N_1\right) \omega\left( \phi \left(N_1\right) \right)
+ 1 \right) H(N_1)^2 \phi'(N_1)^2 + \sum_m \left( 1 + w_m\right) \rho_{m0} \e^{-3 \left( 1 + w_m \right) N_1}\right)\, .
\end{align}
In the FLRW Universe \eqref{FRW}, we can choose $\omega(\phi)=-1$ in \eqref{gfEGB3FLRW} and we obtain,
\begin{align}
\label{phidash}
\phi'=\frac{1}{H} \, .
\end{align}
This allows us to choose $\phi=t$.

We obtain $N=f(\phi)$ and $H= f' (t)$ when the $e$-foldings number $N$ is given by $N=f(t)$.
Then $\xi(\phi)$ and $V(\phi)$ in \eqref{SEGB11}are given as follows,
\begin{align}
\label{SEGB12}
V(\phi) =&\, \frac{3}{\kappa^2} \left( f' (\phi)\right)^2 - 2 \lambda \left( f\left(\phi\right) \right) - \frac{1}{2}
  - \sum_m \rho_{m0} \e^{-3 \left( 1 + w_m \right) f(\phi)} \nonumber \\
&\, - 3 f'(\phi) \e^{ f(\phi)} \int^\phi d\phi_1 \frac{\e^{- f(\phi_1)}}{f'(\phi_1)} \left(\frac{2}{\kappa^2} f'(\phi_1) f''(\phi_1)
+ 2 \lambda \left( f\left(\phi_1\right) \right) + 1
+ \sum_m \left( 1 + w_m\right) \rho_{m0} \e^{-3 \left( 1 + w_m \right) f(\phi_1)}
\right)\, , \\
\label{SEGB13}
\xi(\phi) =&\, \frac{1}{8}\int^\phi d\phi_1
\frac{ \e^{ f(\phi_1)} }{f'(\phi_1)^2} \nonumber \\
&\, \times \int^{\phi_1} d\phi_2 \e^{- f(\phi_2)} \left(\frac{2}{\kappa^2} f'(\phi_2) f''(\phi_2)
+ 2 \lambda \left(f\left(\phi_2\right)\right) + 1
+ \sum_m \left( 1 + w_m\right) \rho_{m0} \e^{-3 \left( 1 + w_m \right) f(\phi_2)}
\right)\, .
\end{align}
Therefore, if we consider the model given by \eqref{SEGB12} and \eqref{SEGB13}, we find one of the solutions of Eq.~\eqref{SEGB3B} as follows,
\begin{align}
\label{SEGB14}
\phi= f^{-1}(N)\quad \left(N= f(\phi)\right)\, ,\quad H = f'\left( f^{-1} \left(N\right) \right) \, .
\end{align}
It should be noted that $\lambda$ can be an arbitrary function of $t$ or $N$ in \eqref{SEGB12} and \eqref{SEGB13}, which allows us to choose $\lambda=0$ in \eqref{SEGB12} and \eqref{SEGB13}, and we obtain,
\begin{align}
\label{SEGB12B}
V(\phi) =&\, \frac{3}{\kappa^2} \left( f' (\phi)\right)^2 - \frac{1}{2}
  - \sum_m \rho_{m0} \e^{-3 \left( 1 + w_m \right) f(\phi)} \nonumber \\
&\, - 3 f'(\phi) \e^{ f(\phi)} \int^\phi d\phi_1 \frac{\e^{- f(\phi_1)}}{f'(\phi_1)} \left(\frac{2}{\kappa^2} f'(\phi_1) f''(\phi_1)
+ 1 + \sum_m \left( 1 + w_m\right) \rho_{m0} \e^{-3 \left( 1 + w_m \right) f(\phi_1)}
\right)\, , \\
\label{SEGB13B}
\xi(\phi) =&\, \frac{1}{8}\int^\phi d\phi_1
\frac{ \e^{ f(\phi_1)} }{f'(\phi_1)^2}
\int^{\phi_1} d\phi_2 \e^{- f(\phi_2)} \left(\frac{2}{\kappa^2} f'(\phi_2) f''(\phi_2)
+ 1 + \sum_m \left( 1 + w_m\right) \rho_{m0} \e^{-3 \left( 1 + w_m \right) f(\phi_2)}
\right)\, .
\end{align}
For the time-evolution of $H$ given by an arbitrary function $g(N)\equiv f'\left( f^{-1} \left(N\right) \right)$ as in \eqref{SEGB14}, we can construct a model realizing the Hubble rate $H$, $H=g(N)$ as in \eqref{GBEDE4} and \eqref{GBEDE5}.

Then, as in \eqref{ex1_4}, we obtain a model generating the inverse phantom crossing by choosing $N=f(\phi)$ and $g(N)\equiv f'\left( f^{-1} \left(N\right) \right)$ so that,
\begin{align}
\label{ex1_3_gf}
\frac{3}{\kappa^2} g\left( f(\phi) \right)^2 = \frac{\Lambda_1}{\e^{\alpha \left(f(\phi)-N_0\right)}
+ \e^{-\alpha \left(f(\phi)-N_0\right)}} + \Lambda_0
+ \sum_m \rho_{m0} \e^{-3 \left( 1 + w_m \right) f(\phi)} \, .
\end{align}
Furthermore, by choosing $N_0\sim -0.4$ and further choosing so that $\Lambda_0$ and $\Lambda_1$ satisfy Eqs.~\eqref{ex1_7} and \eqref{ex1_8}, we may obtain a realistic model.

As in \eqref{itcrssng}, we can also construct the model where the standard phantom crossing and the inverse phantom crossing iterate alternately.

The difference from the scalar--Einstein--Gauss-Bonnet gravity~\eqref{SEGB1} is that there is no fifth force generated by the propagation of the scalar field $\phi$, and there is no fluctuation of the scalar field, which might change the structure formation in the Universe.

\section{Apparent Phantom Crossing}\label{SecV}

Instead of considering the real transition from $w<-1$ to $w>-1$ of the EoS parameter of dark energy, the time-evolution of the dark matter might solve the problem in the DESI observations \cite{Khoury:2025txd}.
Usually, we assume that the energy density of the dark matter redshifts as $a^{-3}$, but if the decrease is slower, it looks as though there could be a phantom crossing in the dark energy sector.

\subsection{Slowly decreasing energy density of  dark matter}

If the dark matter is a non-relativistic particle, the energy density of the dark matter is given by the product of the mass $m_\mathrm{DM}$ and the number density $n_\mathrm{DM}$ of the particle, $\rho_\mathrm{DM} = m_\mathrm{DM} n_\mathrm{DM}$.
When the total number of dark matter particles is conserved, the number density $n_\mathrm{DM}$ is proportional to the inverse of the volume of the Universe $n_\mathrm{DM}\propto V^{-1} \propto a^{-3}$.
If the mass $m_\mathrm{DM}$ is time-dependent $m_\mathrm{DM} (t)$, the energy density of the dark matter is not proportional to $a^{-3}$.

As a concrete example, we may regard the dark matter as a Dirac spinor, whose action is given by,
\begin{align}
\label{DMsp}
S_\mathrm{DM}= \int d^4 x \sqrt{-g} \left\{ i \bar\psi \slashed{D} \psi - m(\phi) \bar\psi \psi \right\} \, .
\end{align}
Here $\slashed{D} \equiv \gamma^\mu D_\mu$ and $D_\mu$ is the covariant derivative including spin connection.
We assume that the mass depends on the scalar field $\phi$ in \eqref{SEGB1} for the scalar-Einstein-Gauss-Bonnet gravity or \eqref{gfEGB4} for the ghost-free $f(\mathcal{G})$ gravity.
The scalar field $\phi$ is identified with the cosmological time $t$ and therefore the mass becomes time-dependent, $m=m(t)$.

In the canonical formulation, the Hamiltonian density $\mathcal{H}$ corresponding to the action \eqref{DMsp} has the following form,
\begin{align}
\label{spH}
\mathcal{H} = \bar\psi \left( - i \gamma^i D_i + m(\phi) \right) \psi\, .
\end{align}
The energy density of the dark matter particle could be given by the quantum or statistical average of the Hamiltonian density,
\begin{align}
\label{sprho}
\rho_\mathrm{DM} = \left< \mathcal{H} \right> = -i \left< \bar\psi \gamma^i \partial_i \psi \right> + m(t) \left< \bar\psi \psi \right> \, .
\end{align}
Because we are considering the non-relativistic particle, the kinetic energy corresponding to the first term in \eqref{sprho}can be very small compared with the rest mass in the second term, so we neglect the first term.
Furthermore, $\left< \bar\psi \psi \right>$ is the number density of dark matter $n_\mathrm{DM}$ and therefore we find,
\begin{align}
\label{sprho2}
\rho_\mathrm{DM} = m(\phi) n_\mathrm{DM} \, .
\end{align}
Because $\phi$ is identified with the cosmological time $t$, if we choose $m(\phi)$ to be an increasing function of $\phi$, $m(\phi)$ becomes larger at late time. Therefore $\rho_\mathrm{DM}$ decreases slower than $a^{-3}$.

If we assume that the energy density of the dark energy is given by subtracting the energy density of dark matter behaving as $\rho_\mathrm{DM} \propto a^{-3}$ from the total energy density, the obtained energy density could not satisfy the energy condition.
If we assume the behavior as in \eqref{sprho2}, each component of the total energy density could not violate any energy condition as observed in \cite{Khoury:2025txd}.

{
You may suspect that the conservation law for the scalar particle might not be violated. 
This is because, due to the Gauss-Bonnet coupling, there is an exchange of energy with the geometry. 
}

\subsection{Energy Conditions}

We now consider the energy conditions, and for the null, weak, strong, and dominant energy conditions, the energy-momentum tensor $T_{\mu\nu}$ satisfies the following inequalities respectively,
\begin{align}
&T_{\mu\nu} k^\mu k^\nu \geq 0\, , \quad T_{\mu\nu} V^\mu V^\nu \geq 0\, , \quad
\left(T_{\mu\nu}-\frac{1}{2}g_{\mu\nu}T \right)V^\mu V^\nu \geq 0\, , \nonumber \\
&T_{\mu\nu} V^\mu V^\nu \geq 0\, , \quad \text{and} \quad T_{\mu\nu} V^\nu \quad \text{is not spacelike} \, .\nonumber
\end{align}
for any null vector $k^\mu$, $g_{\mu\nu}k^\mu k^\nu =0$, and for any time-like vector $V^\mu$, $g_{\mu\nu}V^\mu V^\nu <0$.
In the FLRW Universe~\eqref{FRW}, we consider the perfect fluid whose energy-momentum tensor $T^\mu_{\ \nu}=\text{diag}\left(-\rho , p, p, p\right)$.
In terms of the energy density $\rho$ and the pressure $p$, the above conditions yield,
\begin{eqnarray}
\text{NEC}:&& \quad \rho + p \geq 0 \, , \nonumber \\
\text{WEC}:&& \quad \rho + p \geq 0 \, , \quad \rho \geq 0\, ,\nonumber \\
\text{SEC}:&& \quad \rho + p \geq 0\, , \quad \rho + 3p \geq 0\, ,\nonumber \\
\text{DEC}:&& \quad \rho \geq 0\, , \quad  \rho \geq | p| \, .\nonumber
\end{eqnarray}
By using the EoS parameter $w=\frac{p}{\rho}$, we can further rewrite the above conditions, as follows,
\begin{eqnarray}
\text{NEC}:&& \quad \left( 1+w \right)\rho \geq 0 \, , \nonumber \\
\text{WEC}:&& \quad 1+w \geq 0 \, , \quad \rho \geq 0\, ,\nonumber \\
\text{SEC}:&& \quad \left( 1 + w \right) \rho \geq 0\, , \quad \left( 1+ 3w \right) \rho \geq 0\, ,\nonumber \\
\text{DEC}:&& \quad \rho \geq 0\, , \quad  1 \geq |w| \, .\nonumber
\end{eqnarray}

\subsection{Apparent phantom crossing and energy conditions}

By using the conservation law \eqref{SGBEG9GBN} of the dark energy in terms of $N$, we find,
\begin{align}
\label{SGBEG9GBNp}
p_\mathcal{G} = - \rho_\mathcal{G} - \frac{{\rho_\mathcal{G}}'}{3} \, .
\end{align}
Then the EoS parameter $w_\mathrm{G}$ has the following form,
\begin{align}
\label{SGBEG9GBNpB}
w_\mathcal{G} = - 1 - \frac{{\rho_\mathcal{G}}'}{3\rho_\mathcal{G}} \, .
\end{align}
Then by using \eqref{ex1} and \eqref{ex1_1}, we obtain,
\begin{align}
\label{ex1w}
w_\mathcal{G} = - 1
+ \frac{\frac{\Lambda_1\alpha \left( \e^{\alpha \left(N-N_0\right)} - \e^{-\alpha \left(N-N_0\right)}\right)}
{\left( \e^{\alpha \left(N-N_0\right)} + \e^{-\alpha \left(N-N_0\right)}\right)^2}}
{3\left( \frac{\Lambda_1}{\e^{\alpha \left(N-N_0\right)} + \e^{-\alpha \left(N-N_0\right)}} + \Lambda_0\right)} \, ,
\end{align}
which indicates that { $w_\mathcal{G} > - 1$}, when $N>N_0$ but $w_\mathcal{G} < - 1$, when $N<N_0$.
Therefore, all the energy conditions are violated when $N<N_0$.

For the model \eqref{ex1} with the matter energy density \eqref{ex1_5}, we now investigate the total energy density,
\begin{align}
\label{ttlrh}
\rho = \rho_\mathcal{G} + \rho_m
= \frac{\Lambda_1}{\e^{\alpha \left(N-N_0\right)} + \e^{-\alpha \left(N-N_0\right)}} + \Lambda_0 + \rho_\mathrm{d} \e^{-3N} 
\, ,
\end{align}
We neglect the radiation because we expect the energy density of the radiation could be very small from the epoch before the (inverse) phantom crossing.
The first term on the right-hand side of \eqref{ex1_4} with \eqref{ex1_5}, or the first term in the right-hand side of \eqref{ttlrh}, generates the (inverse) phantom crossing, and the energy conditions can be violated.
The phantom crossing for the total energy density $\rho$ occurs when $\rho'=0$.
We should note $\rho_m'$ is always negative, but $\rho_m''$ is always positive.
On the other hand, Eqs.~\eqref{ex1} and \eqref{ex1_1} indicate that $\rho_\mathcal{G}' > 0$ when $N<N_0$ and $\rho_\mathcal{G}' < 0$ when $N>N_0$ and $\rho_\mathcal{G}'$ takes the maximum when,
\begin{align}
\label{rhGmx}
N=N_\mathrm{max} = N_0 - \frac{1}{2\alpha} \ln \frac{\sqrt{2} + 1}{\sqrt{2}-1}
= N_0 - \frac{1}{\alpha} \ln \left(\sqrt{2} + 1\right)\, .
\end{align}
Then if we choose,
\begin{align}
\label{nopc}
\rho \left( N=N_\mathrm{max} \right)' = \rho_\mathcal{G} \left( N=N_\mathrm{max} \right)'
+ \rho_m \left( N=N_\mathrm{max} \right)' < 0 \, ,
\end{align}
there is no phantom crossing for the total energy density $\rho$.

In the apparent phantom crossing scenario, if we choose $m(\phi)$ in \eqref{sprho2} so that,
\begin{align}
\label{adm}
\rho_\mathrm{DM} = \frac{\Lambda_1}{\e^{\alpha \left(N-N_0\right)} + \e^{-\alpha \left(N-N_0\right)}}
+ \rho_\mathrm{d} \e^{-3N} \, ,
\end{align}
as a simple example, because,
\begin{align}
\label{sdm2}
\rho = \rho_\mathrm{DM} + \Lambda_0 \, , \quad
\rho' = \rho_\mathrm{DM}' \, ,
\end{align}
for the energy density $\rho$ in \eqref{ttlrh}, $\rho_\mathrm{DM}$ does not violate the energy conditions.
Another constituent of the total energy density is only $\Lambda_0$, which can be identified with a cosmological constant.
Therefore, despite an apparent phantom crossing, any constituent of the total energy density does not violate the energy conditions.

We should note that even in this section, we used the behavior of the Hubble rate $H$ in \eqref{ex1_4}, which is used in the previous sections, and it is shown that the behavior could be consistent with the observations of the CMB, BAO, etc.
In the apparent phantom crossing scenario, the total energy density, which determines the Hubble rate $H$, is changed, and each of the components in the energy density does not violate the energy conditions.

It is also possible, as in \eqref{itcrssng}, to construct the model where the standard but apparent phantom crossing and the inverse apparent phantom crossing iterate alternately.

\subsection{Scalar particle scenario}

In this subsection, we consider a scenario of the apparent phantom crossing, which is rather different from the previous scenarios presented in this section.
In this scenario, the apparent (inverse) phantom crossing might be related to the transition from deceleration to acceleration in the expansion of the Universe.

In the case of the scalar--Einstein--Gauss-Bonnet gravity in \eqref{SEGB1}, by the quantization of the scalar field $\phi$, there appear massive particles, which may be candidates for dark matter because the massive particles are regarded a dust.
This situation is different from that of the ghost-free $f(G)$ gravity in \eqref{gfEGB4}, where the scalar field $\phi$ does not propagate and therefore it cannot appear as a particle.
Due to the coupling of the Gauss-Bonnet invariant in the scalar--Einstein--Gauss-Bonnet gravity in \eqref{SEGB1}, the mass $m_\phi$ of the scalar field $\phi$ depends on the curvature as follows,
\begin{align}
\label{massphi}
{m_\phi}^2 \equiv V''(\phi) + \xi''(\phi) \mathcal{G}\, ,
\end{align}
Therefore, the mass might increase at late times, and the decrease in the energy density of the dark matter might be slower than $a^{-3}$, as expected when the dark matter is a dust of particles with a constant mass.

Just for the description, we assume that the effective potential,
\begin{align}
\label{Veff}
V_\mathrm{eff}(\phi )= V(\phi) + \xi(\phi) \mathcal{G}\, ,
\end{align}
has a minimum at $\phi=0$.
Furthermore, we assume
\begin{align}
\label{massphi2}
V(\phi) = V_0 + \frac{1}{2}{m_0}^2 \phi^2 + \mathcal{O}\left(\phi^3\right) \, , \quad
\zeta(\phi) = \frac{1}{2}{m_1}^2 \phi^2 + \mathcal{O}\left(\phi^3\right)\, .
\end{align}
Then the mass of the scalar field is given by,
\begin{align}
\label{massphi3}
{m_\phi}^2 = {m_0}^2 + {m_1}^2 \mathcal{G} + \mathcal{O}\left(\phi\right)\, ,
\end{align}
which depends on the curvature of the spacetime.
In the FLRW Universe~\eqref{FRW}, the Gauss-Bonnet invariant has the following form,
\begin{align}
\label{FLRWG}
\mathcal{G} = 24 H^2 \left( H^2 + \dot H \right)\, .
\end{align}
In the asymptotic de Sitter Universe, we find $\dot H \ll H^2$ and therefore $\mathcal{G}$ is positive.
In the Universe filled with a fluid with the EoS parameter $w$, the Hubble rate $H$ is given by $H\sim \frac{\frac{2}{3(1+w)}}{t}$ and therefore we have,
\begin{align}
\label{Gbhvr}
H^2 + \dot H \sim \left( \frac{4}{9(1+w)^2} - \frac{2}{3(1+w)} \right) \frac{1}{t^2}
= - \frac{2(1 + 3 w)}{9(1+w)^2 t^2} \, .
\end{align}
Then, as long as the Universe undergoes an accelerating expansion with $w<-\frac{1}{3}$, $\mathcal{G}$ is positive, but in the matter-dominated $(w\sim 0)$ or radiation-dominated Universe $\left( w \sim \frac{1}{3} \right)$, $\mathcal{G}$ is negative.
Therefore, the mass in \eqref{massphi3} increases in the epoch near the transition from the decelerated expansion of the Universe to the accelerated expansion of the Universe, although the mass will decrease due to the decrease of $\mathcal{G}$ by the expansion of the Universe.
This indicates that the behavior of the expansion looks as if the phantom regime exists in the epoch of the transition.
We should note that the epoch of the (inverse) phantom crossing was about 5-8 billion years ago, and that time, the transition from deceleration to acceleration in the expansion of the Universe also occurred. Therefore, the transition might be related to the apparent (inverse) phantom crossing.

For simplicity, we neglect $\mathcal{O}\left(\phi\right)$ term in \eqref{massphi3} or $\mathcal{O}\left(\phi^3\right)$ terms in \eqref{massphi2}.
Then, although \eqref{massphi3} tells that $\phi$ becomes a tachyon when $\mathcal{G}$ is negative and large, we only consider the period where ${m_\phi}^2$ is positive.
We may choose $\mathcal{O}\left(\phi^3\right)$ terms in \eqref{massphi2} has another minimum when $\mathcal{G}$ is negative and large, which generates a phase transition, but the tachyon does not appear.
We should note that the particle of $\phi$ should be distinguished from the background field of $\phi$.

If the particle number of the scalar particles corresponding to $\phi$ is conserved, then the number density $n_\phi$ should proportional to $a^{-3}\propto \e^{-3N}$ in the FLRW Universe~\eqref{FRW} and as in \eqref{sprho2}, the energy density $\rho_{\phi\mathrm{DM}}$ of the particles could be given by,
\begin{align}
\label{sprho2DMph} \rho_{\phi\mathrm{DM}} = m_\phi n_\phi
= \rho_{\phi\mathrm{DM}\, 0} \sqrt{ {m_0}^2 + 24 {m_1}^2 H^2 \left( H^2 + \dot H \right)} \e^{-3N} \, ,
\end{align}
where $\rho_{\phi\mathrm{DM}\, 0}$ is a constant.

The first equation in \eqref{SEGB3N}, which corresponds to the first FLRW equation in Einstein's gravity, has the following form,
\begin{align}
\label{SEGB3NphiDM}
0=&\, - \frac{3}{\kappa^2}H^2 + \frac{1}{2}H^2{\phi'}^2 + V_0
+ \frac{1}{2}{m_0}^2 \phi^2 + 24 H^4 {m_1}^2 \phi\phi'
+ \rho_{\phi\mathrm{DM}\, 0} \sqrt{ {m_0}^2 + 24 {m_1}^2 H^2 \left( H^2 + \dot H \right)} \e^{-3N}\nonumber \\
&\, + \sum_m \rho_{m0} \e^{-3 \left( 1 + w_m \right) N} \, .
\end{align}
Here $\sum_m \rho_{m0} \e^{-3 \left( 1 + w_m \right) N}$ expresses the contribution from matter fluids beside the dark matter fluid given by the particles corresponding to $\phi$.
If the scalar field $\phi$ in the background lies in the minimum of the effective potential~\eqref{Veff}, we find $\phi=0$ in the background and we obtain,
\begin{align}
\label{SEGB3NphiDM2}
0=&\, - \frac{3}{\kappa^2}H^2 + V_0
+ \rho_{\phi\mathrm{DM}\, 0} \sqrt{ {m_0}^2 + 24 {m_1}^2 H^2 \left( H^2 + \dot H \right)} \e^{-3N} 
+ \sum_m \rho_{m0} \e^{-3 \left( 1 + w_m \right) N} \, .
\end{align}
In \eqref{SEGB3NphiDM2}, the contributions of $\phi$ in the background vanish, including the Gauss-Bonnet term.
It might be interesting because the Gauss-Bonnet term does not contribute to the equation corresponding to the FLRW equations, but it contributes to the mass of the particles corresponding to $\phi$ via the effective potential~\eqref{Veff}.

In \eqref{SEGB3NphiDM2}, $V_0$ is nothing but the cosmological constant, and we assume $V_0$ is positive.
In the early universe, as in Einstein's gravity, matter is dominant, and the decelerating expansion of the Universe could be realized.
In the late Universe, the cosmological constant $V_0$ becomes dominant as in the standard $\Lambda$CDM scenario, and the accelerating expansion of the Universe started.
Due to the transition from the decelerating expansion to the accelerating expansion, the dark matter given by the particle corresponding to $\phi$, $\rho_{\phi\mathrm{DM}\, 0} \sqrt{ {m_0}^2 + 24 {m_1}^2 H^2 \left( H^2 + \dot H \right)} \e^{-3N}$ generates the apparent phantom crossing.

\section{Summary and Discussions}\label{SecVI}

Although the (inverse) phantom crossing is still not an established phenomenon, we tried to construct realistic models, realizing the phantom crossing.
The Hubble rate in the models are given in \eqref{ex1_4}.
In order to construct the models, we used the scalar--Einstein--Gauss-Bonnet gravity~\eqref{SEGB1}~\cite{Nojiri:2005vv, Nojiri:2006je} and the ghost-free $f(G)$  gravity~\eqref{gfEGB4}~\cite{Nojiri:2018ouv, Nojiri:2022cah}.

We also considered the scenario of the apparent phantom crossing~\cite{Khoury:2025txd}.
First, we proposed a model where the mass of the dark matter depends on the scalar field $\phi$ in \eqref{SEGB1} or \eqref{gfEGB4}.
In the model, any component of the cosmic fluid does not violate the energy conditions, which is different from the phantom scenario.
Also, the energy density of the dark matter decreases more slowly than $a^{-3}$.
We also proposed a new scenario that the particles corresponding to the scalar field $\phi$ in the action \eqref{SEGB1} become dark matter in the framework of the scalar--Einstein--Gauss-Bonnet gravity.
The scenario also suggests that the inverse phantom crossing might be generated by the transition from the decelerating expansion of the Universe to the accelerating expansion.

\end{document}